\documentclass{aa}
\usepackage{psfig}

\begin{document}

%   \thesaurus{12.03.4;	 %cosmology: theory
%	      12.03.1;   %cosmic microwave background
%	      11.09.3;	 %intergalactic medium
%              12.12.1}   %large-scale structure

\title{A halo approach to the evaluation of 
the cross-correlation between the SZ sky and galaxy survey}

\author{Yan Qu and Xiang-Ping Wu}

\offprints{Y. Qu}
\mail{quyan@class2.bao.ac.cn}

\institute{National Astronomical Observatories,
       Chinese Academy of Sciences, Beijing 100012, China}

   \date{Received 00 February, 2003; accepted}

   \titlerunning{SZ-galaxy cross-correlation}

\abstract{
Using a purely analytic approach to gaseous and dark matter halos,
we study the cross-correlation between the Sunyaev-Zel'dovich (SZ) 
sky and galaxy survey under flat sky approximation, 
in an attempt to acquire the redshift information 
of the SZ map. The problem can be greatly simplified when it
is noticed that the signals of the SZ-galaxy correlation
arise only from hot gas and galaxies inside the same 
massive halos (i.e. clusters), and field galaxies 
make almost no contribution to the cross-correlation. 
Under the assumption that both the hot gas and galaxies trace 
the common gravitational potential of dark halos,  
we calculate the expected cross SZ-galaxy power spectra 
for the WMAP/Planck SZ maps and the SDSS galaxy sample
at small scales $100<l<1000$. 
It turns out, however, that it is not presently feasible to
measure such small angular cross power spectra because 
of the high noise levels at $l>400$ with the WMAP/Planck experiments.
Future SZ observations with better angular resolutions and 
sufficiently wide sky coverages
will be needed if this technique is applied for the
statistical measurement of redshift distribution of the SZ sources.

}
 \maketitle

\section{Introduction}

Most of the baryons in the local universe exist in the form of 
warm-hot intergalactic medium with temperature of $T\sim 10^5-10^7$K 
as a result of  gravitationally driven shocks and 
adiabatical compression when they fall into large-scale structures and 
collapsed dark matter halos (e.g. Cen \& Ostriker 1999; Dav\'e et al. 2001).
In the latter case, the very hot baryons gravitationally bound 
in massive halos such as groups and clusters usually manifest themselves by 
strong diffuse X-ray sources in terms of bremmstrahlung emission,
which are directly detectable with current X-ray instruments.
Moreover, the energetic electrons of the hot gas also interact 
the passing cosmic microwave 
background (CMB) photons through the so-called Sunyaev-Zel'dovich (SZ) 
effect,  giving rise to a subtle change in the CMB spectrum.
Current SZ measurements of known clusters at high signal-to-noise 
are now routine (e.g. Carlstrom et al. 2002) and recent detection of 
the excess power relative to primordial CMB anisotropy at arcminute scales 
has been successfully attributed to the statistical signals of 
the thermal SZ effect (Mason et al. 2003; Bond et al. 2003; Komatsu \& Seljak 
 2002).

However, both X-ray and SZ measurements contain no information 
about the redshift distributions of the hot baryons and their
host groups/clusters. Spectroscopic follow-up observations  
should be made to complement our knowledge of the location 
and evolution of the hot baryons and their host halos. It has been
realized that a more practical and powerful approach to extracting 
the distance information from the X-ray and SZ maps 
is perhaps to cross-correlate the X-ray and SZ maps with the 
existing galaxy catalogs or ongoing deep galaxy surveys 
(e.g. Seljak, Burwell \& Pen 2000; Zhang \& Pen 2001;2003;
Zhang, Pen \& Wang 2002).
In particular, the problem can be greatly simplified if we notice
that the signals of the cross-correlation between the SZ (or X-ray) 
map and galaxy survey arise primarily from cluster galaxies, and field
galaxies make almost no contribution to the SZ(or X-ray)-galaxy
correlation. This permits a straightforward calculation of  
the power spectrum of the SZ(or X-ray)-galaxy correlation from 
an analytic model of dark halo abundance, 
i.e., the Press \& Schechter (1974; PS) formalism,
along with a reasonable prescription of galaxy and gas distributions 
inside a given halo of mass $M$ at redshift $z$. For the latter,
one can adopt either the simplest self-similar NFW-like profile 
suggested by  numerical simulations (Navarro, Frenk \& White 1995; NFW),
or the empirically motivated density profiles such as  
the King model or $\beta$-model, in combination with
the halo occupation distribution 
(see Cooray \& Sheth 2003 for a recent review).  
In this paper, we explore the power spectrum of the SZ-galaxy 
cross-correlation based purely on the halo approach. Similar 
technique has been recently applied by Zhang \& Pen (2003)
to the study of the cross-correlation of the soft X-ray 
background and galaxy survey and by Wu \& Xue (2003) 
to the study of the auto-correlation of the soft X-ray background.
In their early work,
Zhang \& Pen (2001) also investigated the SZ-galaxy cross-correlation
using the so-called continuum field model, in which the gas distribution
in dark halos is given by a convolution of the dark matter distribution
with a Gaussian window function, while the matter fluctuation in 
the highly non-linear regime (e.g. clusters) is calculated in terms of  
the hyper-extended perturbation theory.

For the purpose of actual applications, we will calculate 
the expected cross-correlation between 
the SZ maps observed by WMAP/Planck and the galaxy survey 
by the Sloan Digital Sky Survey (SDSS), and assess the feasibility of
extracting the redshift information of the WMAP/Planck SZ maps using
this statistical approach. 
Throughout this paper we adopt a flat cosmological model 
($\Lambda$CDM) with the best fit parameters determined by WMAP:
$\Omega_{\Lambda}=0.73$, $\Omega_M=0.27$, $h=0.71$, 
$\Omega_b h^2=0.0224$, $\sigma_8=0.84$ and $n_s=0.93$

\section{SZ-galaxy cross-correlation}

\subsection{Properties of gaseous and dark matter halos}

The thermal SZ effect along the direction ${\mbox{\boldmath $\theta$}}$ 
due to the hot gas inside a halo can be evaluated following
%1,2
\begin{eqnarray}
\frac{\Delta T({\mbox{\boldmath $\theta$}})}
        {T_{\rm CMB}}&=& g(x)y({\mbox{\boldmath $\theta$}});\\
y({\mbox{\boldmath $\theta$}})&=&
       \int N_e \sigma_T \left(\frac{k_BT}{m_e c^2}\right)\rm d\chi;\\
g(x)&=&\frac{x^2 e^x}{(e^x-1)^2}\left(4-x \coth\frac{x}{2}\right),
\end{eqnarray}
where $N_e$ and $T$ are the number density and temperature of 
electrons,  $x=h_p\nu/k_BT_{\rm CMB}$ is the dimensionless frequency, 
$T_{\rm CMB}$ is the temperature of the present CMB, 
and the integral is performed along the line of sight.
We specify the gas temperature in terms of virial theorem 
%3
\begin{equation}
k_BT=1.39\;{\rm keV}\;f_T
            \left(\frac{M}{10^{15}\;M_{\odot}}\right)^{2/3}
            \left[h^2 E^2(z)\Delta_{\rm c}\right]^{1/3},  
\end{equation}
where $\Delta_{\rm c}$ is the overdensity of dark matter with respect to
the critical value $\rho_{\rm crit}$, which is calculated using a fitting
formula given by Bryan \& Norman (1998) for $f_T=0.8$, and 
$E^2(z)=\Omega_{\rm M}(1+z)^3+\Omega_{\Lambda}$.

For simplicity we assume that both gas and galaxies  
in massive halos trace the dark matter distribution such that
%5,6
\begin{eqnarray}
N_e(r)&=&\frac{f_b}{\mu_e m_p}\Delta\rho_{\rm DM}(r);\\
\Delta N_{\rm gal}(r)&=&
A \Delta \rho_{\rm DM}(r),
\end{eqnarray}
in which $f_b$ is the universal baryon fraction, 
$\mu_e=1.13$ is the mean electron weight, 
$\Delta N_{\rm gal}(r)$ is the number density of galaxies, 
and $A$ is a proportionality constant. 
Indeed, this so-called self-similar model
provides a good approximation for the radial profiles of 
gas and galaxies in dark halos if our focus is not on
the detailed physical processes of the gas and galaxies.
Unlike thermal X-ray emission which is governed by the
central gas content of massive dark halos such as groups
and clusters of galaxies, the thermal SZ effect is generated
by the hot gas distributed over whole halo regions.   
Beyond the core regions  the radial profiles of 
gas and dark matter look roughly very similar in shape. 
In recent years some physically motivated models have been
proposed for the density and temperature of the hot gas
in massive halos (e.g. Babul et al. 2002; Voit et al 2002;
Wu \& Xue 2002; etc.).  One can easily extend our
formalism to inclusion of these gas profiles. Here we adopt the 
universal density profile as suggested by NFW to describe 
the dark matter distribution inside halos, 
$\Delta\rho_{\rm DM}(r)=
\delta_{\rm ch}\rho_{\rm crit}/[(r/r_s)(1+r/r_s)^2]$,
and fix the free parameter using the empirical fitting formula found 
by numerical simulations (Bullock et al. 2001):
$c=[9/(1+z)](M/2.1\times 10^{13} M_{\odot})^{-0.13}$, where 
$c=r_{\rm vir}/r_{\rm s}$ is the so-called concentration parameter. 

In order to determine the constant $A$ in equation (6), 
we need to know the mean galaxy
number for a given halo of mass $M$, i.e. the halo occupation
distribution, $N_{\rm gal}(M)$. We employ 
the best fit analytic formulae of Sheth \& Diaferio (2001) for the 
spiral [$N_{\rm gal,S}(M)$] and elliptical [$N_{\rm gal,E}(M)$]
galaxies based on the GIF simulations (Kauffmannn et al. 1999):
%7
\begin{eqnarray}
N_{\rm gal,S}(M)&=&(M/M_{\rm S})^{\alpha_S}+
                   0.5e^{-4[\log(M/10^{11.75}M_{\sun})]^2}; \nonumber\\
N_{\rm gal,E}(M)&=&(M/M_{\rm E})^{\alpha_E}
                      e^{-(2\times10^{11}M_{\sun}/M)^2}; \nonumber\\
N_{\rm gal}(M)&=&N_{\rm gal,S}(M)+N_{\rm gal,E}(M), 
\end{eqnarray}
where $M_{\rm S}=7\times10^{13}h^{-1}M_{\sun}$, $\alpha_{\rm S}=0.9$,
$M_{\rm E}=3\times10^{12}h^{-1}M_{\sun}$, and $\alpha_{\rm E}=0.75$.
The mean number density of galaxies is thus
%8
\begin{equation}
\overline{N}_{\rm gal}=\int N_{\rm gal}(M) \frac{d^2N}{dMdV} dM,
\end{equation}
where $d^2N/dMdV$ is the comoving number density 
of dark halos (see below). Given the halo occupation
distribution, we are able to normalize the number density profile 
of galaxies within a halo of mass $M$ following 
%9
\begin{equation}
N_{\rm gal}(M)=\int \Delta N_{\rm gal}(r)dV. 
\end{equation}
The integration above is made over whole cluster region out to 
virial radius. This gives the proportionality constant $A$ 
in equation (6) 
%10
\begin{equation}
A=\frac{N_{\rm gal}(M)}
       {4\pi\delta_{\rm ch}\rho_{\rm crit}r_s^3[\ln(1+c)-c/(1+c)]},
\end{equation}
Now, one can easily get the surface number density of galaxies  by
projecting $\Delta N_{\rm gal}(r)$ along the line-of-sight 
${\mbox{\boldmath $\theta$}}$:
%11
\begin{equation}
n_{\rm gal}(M,z,{\mbox{\boldmath $\theta$}})=
D_A^2(z)\int \Delta N_{\rm gal}(r) d\chi,
\end{equation}
in which  $D_A(z)$ is the angular diameter distance to the halo.

\subsection{Power spectra}

The angular cross power spectrum of the SZ-galaxy correlation can be 
separated into the Poisson term $C_l^P$ and 
the clustering term $C_l^C$: 
%12
\begin{eqnarray}
C_l^P&=&g(x)
        \int dz \frac{dV}{dzd\Omega} B(z,z_0) \nonumber\\
& & \int dM\frac{d^2N(M,z)}{dMdV}\vert y_l(M,z)n_l(M,z)\vert,
\end{eqnarray}
and
%13 
\begin{eqnarray}
C_l^{C}&=& g(x)
         \int dz\frac{dV}{dzd\Omega}P(l/D_0,z)B(z,z_0)\nonumber\\
& &\left[\int dM\frac{d^2N(M,z)}{dMdV}b(M,z)y_l(M,z)\right] \nonumber\\
& &\left[\int dM\frac{d^2N(M,z)}{dMdV}b(M,z)n_l\right],
\end{eqnarray}
where $D_0$ is the comoving angular diameter distance to the halo of 
mass $M$ at $z$, and $y_l$ and $n_l$ are the Fourier 
transforms of the Compton $y$-parameter and the surface 
overdensity of galaxies $[n_{\rm gal}(M,z,{\mbox{\boldmath $\theta$}})- 
\overline{n}_{\rm gal}(z_0)]/\overline{n}_{\rm gal}(z_0)$,
respectively, $b(M,z)$ is the bias parameter, for which we use
the analytic prescription of Mo \& White (1996), and
the mean surface number density of galaxies 
$\overline{n}_{\rm gal}(z_0)$ is given by
%14
\begin{equation}
\overline{n}_{\rm gal}(z_0)=\int \overline{N}_{\rm gal}(z)(1+z)^2
                B(z,z_0) D_A^2(z) \frac{d\chi}{dz}dz,
\end{equation} 
where and also in equations (12) and (13) $B(z,z_0)$ denotes the 
redshift distribution function of galaxies. Here we use 
the following approximate expression 
(e.g. Baugh \& Efstathiou 1993; Dodelson et al. 2002):
%15
\begin{equation}
B(z,z_0)=\frac{3z^2}{2(z_0/1.412)^3}\;e^{-(1.412z/z_0)^{3/2}},
\end{equation}
in which $z_0$ is the median redshift of galaxy distribution in
galaxy survey.

Finally, we adopt the PS mass function for the comoving number density 
of dark halos
%16
\begin{equation}
\frac{d^2N}{dMdV}=-\sqrt{\frac{2}{\pi}} \frac{\bar{\rho}}{M} 
    \frac{\delta_{\rm c}(z)}{\sigma^2(M)} 
    \frac{d\sigma(M)}{dM} 
    \exp{\left(-\frac{\delta_c^2(z)}{2\sigma^2(M)} \right)},
\end{equation} 
in which $\delta_c$ is the 
linear over-density of perturbations that collapsed and virialized at 
redshift $z$, and $\sigma(M)$ is the linear theory variance of the 
mass density fluctuation in sphere of mass $M$:
%17
\begin{equation}
\sigma^2(M)=\frac{1}{2\pi^2} 
	\int_0^{\infty} k^2 P(k) {\vert W(kR)\vert}^2 dk,
\end{equation}
and $W(x)=3(\sin x-x\cos x)/x^3$  
is the Fourier representation of the window function.
The power spectrum, $P(k)\propto k^{n_s}T^2(k)$, is normalized by 
the rms fluctuation on an $8$ $h^{-1}$ Mpc scale, $\sigma_8$, and
we take the transfer function $T(k)$ from an adiabatic CDM model given
by Bardeen et al. (1986). 
If we replace  $n_l$ by $g(x)y_l$ and
set $B(z,z_0)=1$ in equations (12) and (13)
we can get the thermal SZ power spectrum, $C^{\rm SZ-SZ}_l$.
Instead, if $g(x)y_l$ is replaced by $n_l$, we will obtain
the galaxy-galaxy correlation,  $C^{\rm gal-gal}_l$. Note that 
for the later case the square of $|n_l|$ in equation (12) should
be $|n_l|$ if $\langle N_{\rm gal}(N_{\rm gal}-1)\rangle<1$ (Seljak 2000).

\section{Application: prediction for the cross-correlation between 
         the MAP/Planck SZ maps and SDSS galaxy survey}

First, we demonstrate in Figure 1 the expected power spectra of 
auto-correlations of the SZ sky and galaxy survey and 
their cross-correlation at frequency $\nu=30$ GHz and 
for all galaxies ($B=1$).  
Note that we have dropped the negative sign in $g(x)$. 
Essentially, the SZ-SZ power spectrum is dominated by 
Poisson distribution over all multipole range from 
$l=100$ (e.g. Komatsu \& Kitayama 1999), 
while clustering of the sources governs the galaxy-galaxy 
correlation at angular scales up to $l\approx 5000$.  
Overall, both of the SZ-SZ and SZ-galaxy power spectra 
exhibit a decline at sufficiently large $l$, regardless of
the fact that the galaxy-galaxy correlation still has much 
power at high $l$ beyond $l=10^4$. 
The latter can be attributed to the pairs of galaxies 
residing within the same, smaller halos at higher 
redshifts which are included in the estimate of 
galaxy-galaxy correlation because of the choice, $B=1$.
Extrapolation of the halo occupation number to high redshifts
may lead to an overestimate of galaxies in small halos. 
Note that unlike the Compton $y$-parameter,  the galaxy
surface number density $n_{\rm gal}(M,z,{\mbox{\boldmath $\theta$}})$
is proportional to $D_A^2(z)$ though both of them have the same 
spatial dependence on the underlying dark matter distribution of 
clusters.  The strong power of the galaxy-galaxy correlation 
at high $l$ can be significantly suppressed when we work with 
a realistic redshift distribution function $B(z,z_0)$ based
on galaxy surveys such as SDSS (see below).  

%%%\placefigure{fig1}

\begin{figure} 
	\psfig{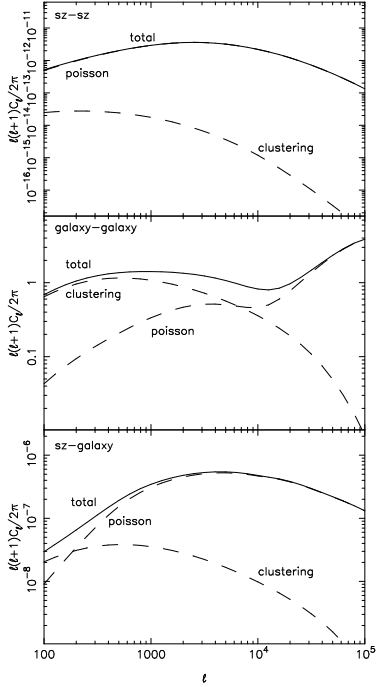}
\caption{The angular power spectra of the SZ-SZ (top panel), 
galaxy-galaxy (middle panel) and SZ-galaxy (lower panel) 
correlations  at frequency $\nu=30$ GHz and for all galaxies. 
Poisson and clustering components are explicitly displayed. }
   \end{figure}

Next, following the investigation of Peiris \& Spergel (2000), we 
carry out our numerical predictions for the galaxy sample 
of SDSS and the SZ sky
(to be) revealed by WMAP/Planck surveyor. We study the feasibility of
extracting the redshift information of the SZ sky from 
their angular cross power spectra. Toward this end, 
we select the SDSS galaxies in different magnitude ranges  
characterized by different values of median redshift $z_0$ in the 
distribution function $B(z,z_0)$ (Dodelson et al. 2002). 
We assume that all the galaxies have photometrically determined 
redshifts to optimize the utilization of the SDSS data. 
This yields a total of
$1.2\times10^{8}$ galaxies over a quarter of the sky 
in the magnitude bin $20<r^*<21$.  
We have also compared the angular power spectra of galaxies 
$C_l^{\rm gal-gal}$ predicted by our halo model with those 
determined by Tegmark et al. (2002) from 
early SDSS data on scales $l<600$ in four different magnitude 
ranges from $r^*=18$ to $22$, and found that the agreement
is good when the bias factors suggested by Tegmark et al. (2002)
are used in different magnitude bins [see Fig.2 of 
Tegmark et al. (2002)].

The working frequencies of both WMAP and Planck are fixed at 
$\nu\approx30$ GHz. Whether or not one is able to effectively 
measure the SZ-SZ, galaxy-galaxy and SZ-galaxy  power spectra 
depends critically on the noise levels. Following the standard
treatment, we estimate the errors on the SZ-SZ power spectrum, 
$C_l^{\rm SZ-SZ}$, and galaxy-galaxy power spectrum, 
$C_l^{\rm gal-gal}$, through
%18,19
\begin{eqnarray}
\Delta C_l^{\rm SZ-SZ}&=&\frac{\sqrt{2}}{\sqrt{(2l+1)\Delta l 
                       f_{\rm sky}^{\rm SZ}}}C_l^{\rm SZ-SZ,tot};\\
\Delta C_l^{\rm gal-gal}&=&\frac{\sqrt{2}}{\sqrt{(2l+1)\Delta l 
                       f_{\rm sky}^{\rm gal}}}C_l^{\rm gal-gal,tot},
\end{eqnarray}
where  $f_{\rm sky}^{\rm SZ}$ and $f_{\rm sky}^{\rm gal}$ are the 
MAP/Planck and SDSS sky coverages, respectively,  
and $\Delta l$ is the bin width
that will be fixed to be $l/4$ in our estimation 
(see Zhang et al. 2002). $C_l^{\rm tot}$ denotes all contributions 
to the measured power spectra that are essentially composed of 
true signals and detector noise, 
$C_l^{\rm tot}=C_l^{\rm true}+C_l^{\rm noise}$, 
provided that foreground contamination is properly removed.   
For the SDSS survey, the shot noise per mode is simply  
%20
\begin{equation}
C_l^{\rm noise}=\frac{1}{n_{\rm gal}}.
\end{equation}
For the SZ sky observed by MAP/Planck, 
the instrumental noise per mode can be modeled by
%21
\begin{equation}
C_l^{\rm noise}=w^{-1}(\nu) e^{\theta^2(\nu)l(l+1)},
\end{equation}
in which we have assumed that the experimental beam is Gaussian 
with width $\theta^2(\nu)$ [FWHM$=(8\ln2)^{1/2}\theta(\nu)$],
and $w^{-1/2}$ denotes the noise level
in CMB temperature fluctuation per pixel. For our working frequency
of $\nu\approx30$ GHz, we choose these parameters in terms of the 
tabulated values given by Cooray \& Hu (2000). 
In a similar way, we can estimate the error for the SZ-galaxy 
power spectrum from
%22
\begin{eqnarray}
\Delta C_l^{\rm SZ-gal}&=&\frac{\sqrt{2}}{\sqrt{(2l+1)\Delta l f_{\rm sky}}}
                   \left[\left(C_l^{\rm SZ-gal}\right)^2 \right. \nonumber\\
                       & & +\left.C_l^{\rm SZ-SZ,tot}
                          C_l^{\rm gal-gal,tot}\right]^{1/2}.
\end{eqnarray}

We show in Figure 2 the expected power spectra of the 
auto-correlations of the WMAP/Planck SZ maps and the SDSS galaxy survey 
and their cross-correlations for $z_0=0.33$, corresponding to
$20<r^*<21$. 
While in principle the galaxy power spectrum can be well constructed 
with the SDSS survey over a broad range of multipoles, 
the measurements of the SZ power spectra beyond $\l\sim$ a few hundreds 
are hampered by the poor angular resolutions of the WMAP/Planck 
experiments, although the result of Planck is slightly better
[see Bennett et al. (2003) for the first year WMAP result]. 
Recall that on the observed CMB spectrum the SZ signals appear
to be dominant only at $>2000$ unless the primary CMB can be 
accurately removed. Consequently, it may become 
hopeless to measure the cross-correlation between the WMAP/Planck SZ sky
and the SDSS galaxy survey. Yet, another possibility is to search for
the cross-correlation signals at very large angular scales generated 
by superclusters (Hern\'andez-Monteagudo \& Rubin\~o-Mart\'in 2003)
or large-scale matter inhomogeneities at low redshifts (e.g. $z<0.1$) 
(Afshordi, Loh \& Strauss 2004). Indeed, the noise appears to be relatively 
smaller at low redshifts and large angular scales (see Figure 3),
which facilitates the detection of the cross-correlation of the 
WMAP/Planck SZ maps with galaxy surveys such as SDSS and 2MASS,
although in this case our primary goal of extracting redshift 
information of the SZ map from the cross-correlation signals 
becomes less interesting.

%\placefigure{fig2}

\begin{figure} 
	\psfig{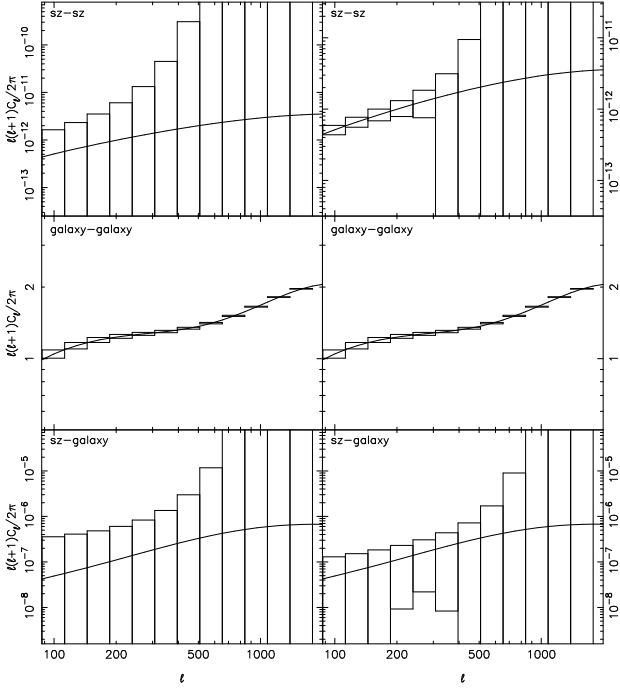}
\caption{Expected angular power spectra of the SZ-SZ (top panel), 
the SDSS galaxy-galaxy (middle panel) and the SZ - SDSS galaxy 
(lower panel) correlations  at frequency $\nu=30$ GHz and 
for $z_0=0.33$.  Bin width is chosen as $\Delta l=l/4$. Left and right
panels correspond to the WMAP and Planck results, respectively.}
   \end{figure}

Taking the Planck experiment as an example, we have shown in Figure 3 
the SZ-galaxy cross-correlations for a set of median redshifts 
of the SDSS galaxies. Despite the
large errors at high $l$ beyond $\sim400$, it demonstrates the actual 
procedures that one needs to perform in order to extract the redshift 
information of the SZ map from these cross-correlations.

%\placefigure{fig3}

    \begin{figure}
	\psfig{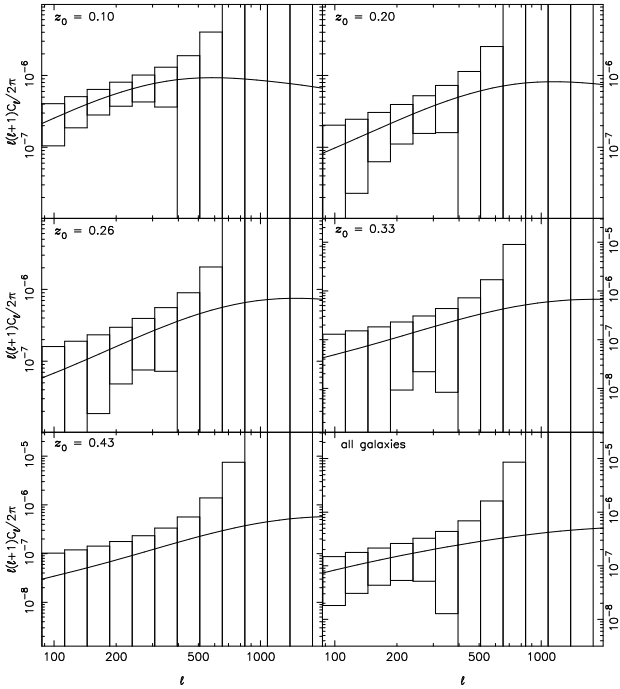}
        \caption{Dependence of the Planck SZ-galaxy power spectra 
on redshifts of the SDSS galaxies.}
   \end{figure}

Finally,  we calculate the correlation coefficient between 
the SZ map and galaxy survey following
%23
\begin{equation}
r_{\rm corr}(l) = \frac{C_l^{\rm SZ-gal}}{\sqrt{C_l^{\rm SZ-SZ} 
               C_l^{\rm gal-gal}}}.
\end{equation}
Recall that $r_{\rm corr}(l)$ acts as a quantitative 
indicator of the significance 
of the SZ-galaxy correlation.  The associated error can be 
estimated by 
%24
\begin{eqnarray}
(\Delta r)^2&=& r_{\rm corr}^2(l)\left\{
           \left(\frac{\Delta C_l^{\rm SZ-gal}}{ C_l^{\rm SZ-gal}}\right)^2
          +\frac{1}{2(2l+1)f_{\rm sky}\Delta l} \right.\nonumber \\
        & &\times\left(\frac{C_l^{\rm SZ-SZ,tot}}{C_l^{\rm SZ-SZ}}+
           \frac{C_l^{\rm gal-gal,tot}}{C_l^{\rm gal-gal}}\right)\nonumber\\
        & & \left[\frac{C_l^{\rm SZ-SZ,tot}}{C_l^{\rm SZ-SZ}}+
                  \frac{C_l^{\rm gal-gal,tot}}{C_l^{\rm gal-gal}} \right. 
                                                   \nonumber\\
        & &   \left.\left.
                -4\left(1+\frac{C_l^{\rm SZ-SZ,tot}C_l^{\rm gal-gal,tot}}
                        {(C_l^{\rm SZ-gal})^2}\right)^{1/2}\right]\right\}.
\end{eqnarray}
We display in Figure 4 the resulting correlation coefficients 
in the same redshift ranges as in Figure 3 for the Planck SZ-SDSS galaxy 
correlation. Similarly to the power spectra in  Fig.3,  instrumental 
noise becomes to be dominant at both lower and higher $l$ for the 
correlation coefficients. Peak location of $r_{\rm corr}$ increases
with the median redshift of SDSS galaxy distribution because 
inclusion of more distant galaxies in the estimate 
of the SZ-galaxy correlation raises the power at smaller angular scales.  
Overall, the correlation efficient has a value of 
$r_{\rm corr}\approx0.2$ at $100<l<1000$,  indicating that 
the two phenomena are only moderately correlated with each other.
This arises from the fact that only can galaxies in clusters contribute 
to the  SZ-galaxy correlation and field galaxies selected in SDSS
sample have almost no effect on the cross correlation between the SZ map
and galaxy survey. 
Zhang \& Pen (2001) made a similar analysis to maximize 
$r_{\rm corr}$ by properly picking the redshift distribution
function of the SDSS galaxies, which can increase the correlation 
efficient to 
$r_{\rm corr}^{\rm max}\approx0.7$ at a smaller scale of $10^3<l<10^4$. 

%\placefigure{fig5}

    \begin{figure} 
	\psfig{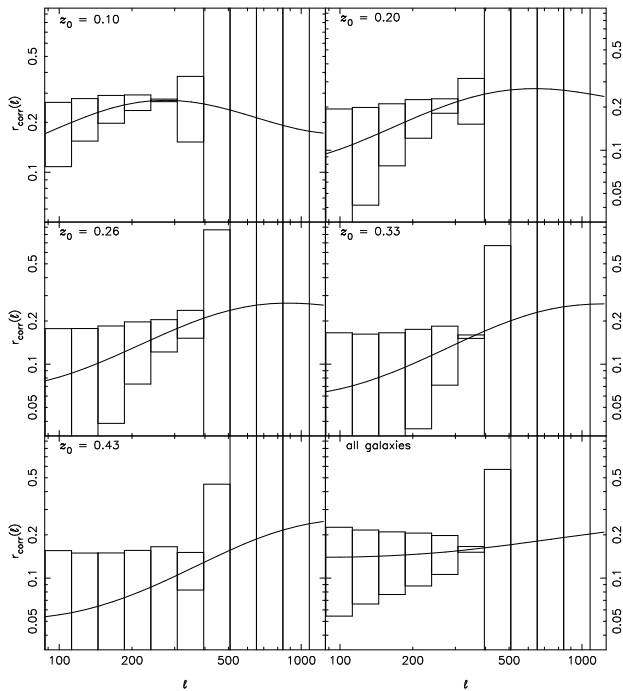}
        \caption{The correlation coefficients for the Planck SZ map 
and SDSS galaxy survey in different redshift ranges. Bin width is
taken to be $\Delta l=l/4$.}
   \end{figure}

\section{Discussion and conclusions}

Cross-correlation between the SZ map and galaxy survey 
will reveal, in a statistical manner, important information 
on the redshift distribution of hot baryons in the universe,
which requires no spectroscopic follow-up observation of
individual sources. Recall that the major advantage of 
measurement of the SZ power spectrum over the SZ cluster survey   
is that one can acquire weak SZ signals at high statistical 
significance level without the need for resolving individual clusters.
Indeed, time-consuming spectroscopic follow-up observations 
of the SZ sources may eventually throw a shadow 
on the effective applications of the
SZ power spectrum if redshift distribution is ultimately 
concerned.  Cross-correlation between the SZ map and 
galaxy survey provides a simple approach which is just 
suited for the problem. In particular, theoretical prediction of 
the SZ-galaxy power spectrum can be greatly simplified if 
one notices that the signals of the SZ-galaxy cross-correlation 
are primarily produced by the hot gas and galaxies within the same 
clusters. Namely, field galaxies make almost no contribution to
the SZ-galaxy cross-correlation. 

Assuming that both the intracluster gas and cluster galaxies follow
the same dark matter distribution with a functional form of the NFW-like 
profile and the abundance of dark halos is described by the PS mass 
function, we have predicted the SZ-galaxy angular power spectrum.
As it is expected, the power spectrum indeed shows a moderately  
strong correlation indicated by an overall correlation coefficient 
of $r_{\rm corr}\approx 0.2$ at $100<1<1000$. 
However,  applying our algorithm to 
the SDSS galaxy survey and MAP/Planck SZ maps yields an unpleasant
result: It is unlikely that one can acquire meaningful information 
about the redshift distribution of the MAP/Planck SZ maps because 
the SZ signals are significantly below the noise levels 
in our interested range of multipoles, $l>1000$, 
although the SDSS galaxy sample
should be just suited for such a purpose.  
One possibility is to work at large angular scales $l<100$, which 
is related to the SZ effect produced by nearby superclusters and
clustering of galaxies (Hern\'andez-Monteagudo \& Rubin\~o-Mart\'in 2003;
Afshordi et al. 2004). Another application 
is to utilize the CMB data at very small angular scales beyond $l=2000$
to be obtained by many ongoing/upcoming  experiments such as ACBAR, 
AMiBA, BIMA, CBI, etc. In this case, one needs instead to deal with 
the small-sky coverage problem. Indeed, a similar work should be done 
on the feasibility of cross-correlating these small-sky 
coverage CMB data with  existing galaxy catalogs. We conclude that 
the cross-correlation between the SZ map and galaxy survey
can in principle yield valuable information about the redshift
distribution of the host baryons in the universe. However, 
the actual applications of this technique to real observations may 
not be possible until high angular resolution and sensitivity
SZ power spectrum over a wide sky coverage is achieved.

\begin{acknowledgements}
Constructive suggestions by an anonymous referee are gratefully acknowledged. 
This work was supported by the National Science Foundation of China, 
under Grant No. 10233040, and the Ministry of Science and Technology 
of China, under Grant No. NKBRSF G19990754.
\end{acknowledgements}

\end{document}